\documentclass[aps,nofootinbib,twocolumn,prd,showpacs,showkeys,preprintnumbers]{revtex4-1}

\usepackage[caption=false]{subfig}
\usepackage{graphicx}
\usepackage{amsmath}
\usepackage{amsfonts}
\usepackage{amssymb}
\usepackage{float}
\usepackage[usenames]{color}
\usepackage{bm}
\usepackage{mathrsfs}
\usepackage{epstopdf}
\usepackage{url}
\usepackage{footnote}
\usepackage{textcomp}

\makeatletter
\newcommand*{\rom}[1]{\expandafter\@slowromancap\romannumeral #1@}
\makeatother

\begin{document}

\title{Is Eddington-Born-Infeld theory really free of cosmological singularities?}
\author{Mariam Bouhmadi-L\'{o}pez$^{1,2}$}
\email{mariam.bouhmadi@ist.utl.pt}
\author{Che-Yu Chen $^{3,4}$}
\email{b97202056@ntu.edu.tw}
\author{Pisin Chen $^{3,4,5,6}$}
\email{chen@slac.stanford.edu}
\date{\today}

\affiliation{
${}^1$Department of Theoretical Physics, University of the Basque Country
UPV/EHU, P.O. Box 644, 48080 Bilbao, Spain\\
${}^2$IKERBASQUE, Basque Foundation for Science, 48011, Bilbao, Spain\\
${}^3$Department of Physics, National Taiwan University, Taipei, Taiwan 10617\\
${}^4$LeCosPA, National Taiwan University, Taipei, Taiwan 10617\\
${}^5$Graduate Institute of Astrophysics, National Taiwan University, Taipei, Taiwan 10617\\
${}^6$Kavli Institute for Particle Astrophysics and Cosmology, SLAC National Accelerator Laboratory, Stanford University, Stanford, CA 94305, U.S.A.
}

\begin{abstract}
The Eddington-inspired-Born-Infeld (EiBI) theory has been recently resurrected. Such a theory is characterized by being equivalent to Einstein theory in vacuum but differing from it in the presence of matter. One of the virtues of the theory is to avoid the Big Bang singularity for a radiation filled universe. In this paper, we analyze singularity avoidance in this kind of model. More precisely, we analyze the behavior of a homogeneous and isotropic universe filled with phantom energy in addition to the dark and baryonic matter. Unlike the Big Bang singularity that can be avoided in this kind of model through a bounce or a loitering effect on the physical metric, we find that the Big Rip singularity is unavoidable in the EiBI phantom model even though it can be postponed towards a slightly further future cosmic time as compared with the same singularity in other models based on the standard general relativity and with the same matter content described above.
\end{abstract}

\maketitle
Einstein theory of general relativity (GR) is an extremely successful theory for nearly a century \cite{gravitation}. Despite of all its advantages, it is expected to break down at some point at very high energies, for example in the past evolution of the Universe where the theory predicts a Big Bang singularity \cite{largescale} and the laws of physics cease to be valid. This is one of the motivations for looking for possible extension of GR. In addition, it is hoped that modified theories of general relativity, while preserving the great achievements of GR, would shed some light over the unknown fundamental nature of dark energy or whatsoever stuff that drives the present accelerating expansion of the Universe (see for example Ref.~\cite{reviewmodifiedGR}). 

There have been many proposals for alternative theories of GR almost as old as the theory itself. One of the oldest was proposed by Eddington \cite{Eddington}, where the connection rather than the metric plays the role of the fundamental field of the theory. The gravitational action proposed by Eddington back in 1924 \cite{Eddington} is equivalent to Einstein theory of GR in vacuum. One of the weak points of the theory is that it does not incorporate matter. Recently, an Eddington-inspired-Born-Infeld theory (EiBI)  has been proposed in Ref.~\cite{Banados:2010ix} where matter fields are incorporated into the Lagrangian formulation. More importantly, it turns out that this theory avoids the Big Bang singularity that would face a radiation-dominated universe in standard GR \cite{Banados:2010ix}. The \textit{apparent} fulfillment of the energy conditions in EiBI theory was considered in Ref.~\cite{Delsate:2012ky}, where the adjective \textit{apparent} refers to quantities defined with respect to a metric compatible with the connection that defines the theory. Their analysis leads to a sufficient condition for singularity avoidance. Besides, the gravitational collapse of noninteracting particles, i.e., dust or equivalently pressureless matter, does not lead to singular states in the nonrelativistic limit (Newtonian regime) \cite{Pani:2011mg} (see also \cite{Pani:2012qb}).
This theory has also been studied as an alternative scenario to the inflationary paradigm \cite{Avelino:2012ue}. Furthermore, possible constraints on the parameter characterizing the theory have been obtained using solar models \cite{Casanellas:2011kf}, neutron stars \cite{Avelino:2012ge}, and nuclear physics \cite{Avelino:2012qe}. It has also been shown that such avoidance of Big Bang singularity is more general and not limited to the radiation-dominated universe \cite{Scargill:2012kg}. Despite of all the virtues of the EiBI theory, a cosmological tensor instability in this model was found in Ref.~\cite{EscamillaRivera:2012vz}. In addition, this theory behaves similarly to the Palatini f(R) gravity and shares the same pathologies, such as curvature singularities at the surface of polytropic stars and some unacceptable phenomenology \cite{Pani:2012qd}.

In this letter, we ask the simple questions: Is EiBI theory really free of cosmological singularities? In particular, is the theory free of dark energy related singularities? In Ref.~\cite{Delsate:2012ky}, it was shown that if the null energy condition is fulfilled, then the \textit{apparent} null energy condition is satisfied. It turns out that the null energy conditions are not always fulfilled; a clear example of it is a super-inflationary phase within GR. Moreover, in recent years a new singularity named the Big Rip has been identified where the null energy condition is in fact not fulfilled and the Universe is ripped apart: the Hubble rate and its cosmic derivative approach infinity in a finite cosmic time \cite{Starobinsky:1999yw,Caldwell:2003vq}. Can such singularity be avoidable in the theory proposed in Ref.~\cite{Banados:2010ix}? This question is even more pertinent in the aftermath of the release of WMAP9 data, which hints on the possibility of a phantom energy component in the Universe more pronouncedly than that deduced from the WMAP7 data \cite{Komatsu:2010fb}. The analysis of the possible occurrence of a Big Rip in the future of the Universe is therefore timely.

Our starting point is the gravitational action with the metric $g_{\mu\nu}$ and connection $\Gamma^{\alpha}_{\mu\nu}$ recently proposed in 
\cite{Banados:2010ix}:
\begin{eqnarray}
\mathcal{S}_{\textrm{EiBI}}(g,\Gamma,\Psi)&=&\frac{2}{\kappa}\int d^4x\left[\sqrt{|g_{\mu\nu}+\kappa R_{\mu\nu}(\Gamma)|}-\lambda\sqrt{g}\right] \nonumber\\ &\,&+\,\mathcal{S}_\textrm{m} (g,\Gamma,\Psi),
\label{action}
\end{eqnarray} 
where $R_{\mu\nu}(\Gamma)$ stands for the symmetric part of the Ricci tensor and, as indicated in Eq.~(\ref{action}), is constructed from the connection $\Gamma$. We consider the action under the Palatini formalism, i.e., the connection $\Gamma^{\alpha}_{\mu\nu}$ is {\it not} the Levi-Civita connection of the metric $g_{\mu\nu}$. The parameter $\kappa$ is a constant with inverse dimensions to that of the cosmological constant (in this letter, we will work with Planck units $8\pi{G}=1$ and set the speed of light to $c=1$), $\lambda$ is a dimensionless constant and $\mathcal{S}_\textrm{m} (g,\Gamma,\Psi)$ stands for the matter Lagrangian. 
This Lagrangian has two well defined limits: (i) when $|\kappa R|$ is very large, we recover Eddington's theory and (ii) when $|\kappa R|$ is small, we obtain the Hilbert-Einstein action with an effective cosmological constant $\Lambda=(\lambda-1)/\kappa$ \cite{Banados:2010ix}. A solution of the action in Eq.(\ref{action}) can be characterized by two different Ricci tensors: 
$R_{\mu\nu}(\Gamma)$ as presented in Eq.(\ref{action}) and $R_{\mu\nu}(g)$ constructed from the metric $g$. There are in addition three ways of defining the scalar curvature. These are: $g^{\mu\nu}R_{\mu\nu}(g)$, $g^{\mu\nu}R_{\mu\nu}(\Gamma)$ and $R(\Gamma)$. The third one is derived from the contraction between $R_{\mu\nu}(\Gamma)$ and the metric compatible with the connection  $\Gamma$. Therefore whenever one refers to singularity avoidance, one must specify the scalar curvature(s).

The equations of motion are obtained by varying the action, $\mathcal{S}_{\textrm{EiBI}}$, with respect to the metric and the connection. The energy-momentum tensor is conserved in this theory. Thus, for a Friedmann-Lema\^{\i}tre-Robertson-Walker (FLRW)
universe filled with a perfect fluid with energy density $\rho$ and pressure $p$, we obtain the familiar relation
\begin{equation}
\dot\rho+3 H(\rho+p)=0.
\label{conservation equation}
\end{equation}
After the variation of the action (\ref{action}) and combining it with the conservation equation Eq.(\ref{conservation equation}), we arrive at a modified Friedmann equation for a universe with scale factor $a$, which is filled with a perfect fluid with energy density $\rho$ and pressure $p$ \cite{Banados:2010ix}:
\begin{equation}
H^2=\frac{2}{3}\frac{G}{F^2},
\label{field equation}
\end{equation}
where $\rho_t=\rho+\Lambda$, $p_t=p-\Lambda$, $\Lambda=(\lambda-1)/\kappa$,
\begin{eqnarray}
F&=&2-\frac{3\kappa(\rho_t+p_t)\left[1-\kappa p_t -\frac{dp_t}{d\rho_t}(1+\kappa\rho_t)\right]}{2(1+\kappa \rho_t)(1-\kappa p_t)}, \label{F}\\
G&=&\frac{1}{\kappa}\left(1+2U-3 \frac{U}{V}\right), \label{G}\\
U&=&(1-\kappa p_t)^{\frac32}(1+\kappa \rho_t)^{-\frac{1}{2}},V=U^\frac13(1+\kappa\rho_t)^\frac23.\nonumber
\end{eqnarray}
Consequently, for a universe whose matter content is dominated by a single component with its equation of state (EOS) $p=w\rho$, and in the absence of a cosmological constant, i.e., $\Lambda=0$, its evolution is governed by
\begin{eqnarray}
H^2&=&\frac{8}{3\kappa}\Big[(1+3w)\bar\rho-2+2\sqrt{(1+\bar\rho)(1-w\bar\rho)^3}\Big]\nonumber\\
&\times&\frac{(1+\bar\rho)(1-w\bar\rho)^2}{(4+(1-w)(1-3w)\bar\rho+2w(1+3w)\bar\rho^2)^2},
\label{field equation w}
\end{eqnarray}
where $\bar\rho=\kappa\rho$ \cite{Scargill:2012kg}. It can be easily verified that the Big Bang singularity can be avoided in this theory for a radiation dominant universe; i.e., $w=1/3$. More specifically, the Universe either bounces in the past for the case of $\kappa<0$, or has a loitering behavior in the infinite past for the case of $\kappa>0$ \cite{Banados:2010ix}. Despite that the Big Bang singularity is avoided with respect to the metric $g$; i.e., the Hubble rate, its cosmic time derivative, scalar curvature $g^{\mu\nu}R_{\mu\nu}(g)$, and Ricci curvature $R_{\mu\nu}(g)$ are all finite, nevertheless the scalar curvature of the metric compatible with the connection $\Gamma$, i.e., $R(\Gamma)$, still diverges when the scale factor approaches the minimum (see TABLE \ref{curvature}).

A natural question, inspired by this finding, is how general is the singularity avoidance in EiBI as compared with GR. In particular, can EiBI cure or smoothen the Big Rip singularity? Such a singularity is expected in GR for a phantom energy dominated universe with a constant equation of state. In order to address this question, we focus on the late-time evolution of a FLRW universe filled with phantom energy ($w\lesssim-1$ and is constant) in addition to the dark and baryonic matter. The matter content reads
\begin{eqnarray}
\rho_t&=&\rho_m+\rho_w=\rho_{m_0}a^{-3}+\rho_{w_0}a^{-3(1+w)},\nonumber\\
p_t&=&p_w=w\rho_{w_0}a^{-3(1+w)},
\label{constituents}
\end{eqnarray}
where $\rho_m$ and $\rho_w$ are the density of matter and dark energy, respectively. 

This model contains four parameters, $\rho_{m_0}$, $\rho_{w_0}$, $w$, and $\kappa$, but only three are independent because of the cosmological constraint obtained by evaluating the Friedmann equation at the present time, which reads\footnote{The parameters $\Omega_m$ and $\Omega_w$ are defined in the standard way, i.e., $\Omega_m\equiv\rho_{m_0}/\rho_c$ and $\Omega_w\equiv\rho_{w_0}/\rho_c$, where $\rho_c$ is the critical density. In addition, $\Omega_\kappa\equiv\kappa\rho_c=3{H_0}^2\kappa$, where $H_0$ is the current Hubble parameter.}:
\begin{equation}
\Omega_\kappa=f(\Omega_\kappa, \Omega_m, \Omega_w, w)=\frac{2W}{X^2},
\label{constraint}
\end{equation}
where
\begin{eqnarray}
W&=&1+2\frac{A^{\frac{3}{2}}}{B^{\frac{1}{2}}}-3\frac{A}{B},\nonumber\\
X&=&2-\frac{3\Omega_\kappa C(A-Y)}{2AB},\nonumber\\
Y&=&\frac{\Omega_w[1+w(1+w)\Omega_\kappa(\Omega_m+\Omega_w)]}{C},
\end{eqnarray}
and $A=1-w\Omega_\kappa\Omega_w$, $B=1+\Omega_\kappa(\Omega_m+\Omega_w)$, $C=\Omega_m+(1+w)\Omega_w$, respectively.

At small $\Omega_\kappa$, $f(\Omega_\kappa, \Omega_m, \Omega_w, w)$ can be written as $f(\Omega_\kappa, \Omega_m, \Omega_w, w)\approx(\Omega_m+\Omega_w)\Omega_\kappa+O({\Omega_\kappa}^2)$, which confirms again that EiBI reduces to GR for vanishing $\Omega_\kappa$.

Since EiBI theory contains a new parameter, $\kappa$, as compared with GR, we shall first put some constraints in it before proceeding further. The estimation will be based on three points as follows: (i) We expand the right-hand-side of the cosmological constraint, 
Eq.(\ref{constraint}), up to the second order in $\Omega_\kappa$ and obtain the solutions:
\begin{eqnarray}
\Omega_\kappa&=&0,\nonumber\\
\Omega_\kappa&=&\frac{8}{3K}[1-(\Omega_w+\Omega_m)],
\label{twosolution}
\end{eqnarray}
where
\begin{eqnarray}
K&=&\Omega_m^2-2\Omega_m\Omega_w-3\Omega_w^2\nonumber\\
&+&2\Omega_m\Omega_ww+2\Omega_w^2w+\Omega_w^2w^2.
\label{K}
\end{eqnarray}
This approximation is used here as a way to simplify the presentation of our results and does not affect the conclusions of our paper. 
(ii) We assume that the model conforms with the $w$CDM scenario at present so as to be able to explain the current acceleration of the Universe. We may therefore assume that $0.267\le\Omega_m\le0.287$, $0.713\le\Omega_w\le0.733$, and $-1.147\le w\le-1.021$ \cite{Komatsu:2010fb}. These cosmological constraints imply that $K<0$ (cf. Eq.~(\ref{K})).
(iii) We restrict our model to a positive $\Omega_\kappa$, i.e., $\kappa>0$, in order to avoid the imaginary effective sound speed instabilities usually present in EiBI theory \cite{Avelino:2012ge}. As a consequence, $\Omega_\kappa$ vanishes whenever $\Omega_m+\Omega_w\leq1$. (See Eq.~(\ref{twosolution}) and (ii).)

In summary, under the above conditions we can conclude the following: (i) $\Omega_\kappa=0$ whenever $\Omega_m+\Omega_w\leq1$, where we recover GR. (ii) For $\Omega_m+\Omega_w>1$, we may consider the second solution in Eq.~(\ref{twosolution}), which is positive in this case. In the latter case, we will assume that $\Omega_\kappa$ is small, i.e., the deviation of EiBI theory from GR is small, so that $\Omega_m+\Omega_w\gtrsim1$ is in agreement with the observational data \cite{Komatsu:2010fb}. (iii) One can always find a suitable value for $\Omega_\kappa$, or $\kappa$, to fit a specific set of parameters $\Omega_m$, $\Omega_w$, and $w$.

We now investigate the asymptotic behavior of the Universe within this framework. This amounts to determining the Hubble parameter, $H$, and its cosmic time derivative, $\dot{H}$, at large scale factors. From here on we set $w=-1-\epsilon$, where $\epsilon$ is positive. As the dark energy corresponds to the phantom matter in our setup and the Universe is expanding, the conservation of such an energy density implies a growth of $\rho_w$ (see Eq.~(\ref{conservation equation})), unlike the baryonic and dark matter, which would quickly become negligible as compared with $\rho_w$. We therefore neglect $\rho_m$ in our estimation of $H$ and $\dot{H}$.

Under the above assumptions, ($1\ll\kappa\rho_t\approx\kappa\rho_w$), we obtain the asymptotic behavior of the Hubble parameter $H$ given in Eq.~(\ref{field equation}) and the cosmic time derivative, $\dot{H}$, by simply combining Eq.~(\ref{conservation equation}) and Eq.~(\ref{field equation}):
\begin{eqnarray}
H^2&\approx&\frac{4\sqrt{(1+\epsilon)^3}}{3(2+3\epsilon)^2}\rho_t,\nonumber\\
\dot{H}&\approx&\frac{2\sqrt{(1+\epsilon)^3}}{(2+3\epsilon)^2}\epsilon\rho_t.
\end{eqnarray} 
The above results correspond to the dominant terms in the expansion of $\kappa H^2$ and $\kappa\dot{H}$ as functions of $\kappa\rho_t\approx\kappa\rho_w$. As can be seen, $H$ and $\dot{H}$ will blow up when $\kappa\rho_w$ diverges at an infinite radius of the Universe.

We can also prove that a phantom energy dominated EiBI universe has a well defined $H^2$ for any value of $\rho_w$. In fact, the square of the Hubble parameter in Eq.~(\ref{field equation w}), for $\bar\rho=\kappa\rho_w$, is positive-definite and it vanishes only when $\rho_w=0$.

So far we have shown that the total density $\rho_t$, the total pressure $p_t$, $H^2$, and $\dot{H}$ will all diverge when the scale factor goes to infinity. Our next step is to confirm the existence of the Big Rip singularity at some finite cosmic time. The cosmic time can be evaluated directly from the integral
\begin{equation}
H_0(t_{\textrm{sing}}-t_0)=\int_{-1}^{0}\frac{dz}{(1+z)E(z)},
\label{int}
\end{equation}
where $z$ is the redshift parameter, $E(z)=H/H_0$,  $t_{\textrm{sing}}$ and $t_0$ are the cosmic time the singularity takes place and the present time, respectively. For the relativistic Friedmann equation, we have $E_{\textrm{cl}}(z)=[\Omega_m{(z+1)}^3+\Omega_w{(z+1)}^{-3\epsilon}]^{1/2}$
and for the EiBI modified theory, we can use Eq.~(\ref{field equation}) with the matter content given in Eq.~(\ref{constituents}) to derive $E_{\textrm{EiBI}}(z)$, where $E_{\textrm{EiBI}}(z)=H_{\textrm{EiBI}}(z)/H_0$. We show in TABLE~\ref{tabletime} the results of our numerical integration based on Mathematica 7, where we have assumed $\Omega_m=0.287$ and $\Omega_w=0.733$  \cite{Komatsu:2010fb}, and used the constraint Eq.(\ref{constraint}) to find the corresponding $\Omega_\kappa$. We choose those limiting values of $\Omega_m$, $\Omega_w$ to enhance the possible effects of the EiBI model; i.e., we choose observational values that maximize the inequality $\Omega_m+\Omega_w\gtrsim1$. We see that the cosmic time derived from the EiBI theory is finite and of the value of ten times the current age of the Universe, which implies that this theory is not able to remove the Big Rip singularity occurring in GR, even though this singularity can be slightly pushed towards a future time as compared with GR.

Our results indicate that the scalar curvature constructed from the physical metric $g_{\mu\nu}$ will blow up at the Big Rip. It can be shown that $R_{\mu\nu}(\Gamma)$ and $g^{\mu\nu}R_{\mu\nu}(\Gamma)$ also blow up at $t_{\textrm{sing}}$ where $a$ diverges, whereas $R(\Gamma)$ remains finite. Specifically, $R_{00}(\Gamma)=(1-U)/\kappa\rightarrow-\infty$ and $R_{ij}(\Gamma)=[a^2(V-1)\delta_{ij}]/\kappa\rightarrow\infty$, while $R(\Gamma)=(U-1)/U\kappa+3(V-1)/V\kappa\rightarrow 4/\kappa$, as $a\rightarrow\infty$.

An interesting model for a modified theory of gravity was suggested in Ref.~\cite{Banados:2010ix}. It was shown that in this model the Big Bang singularity for a radiation-filled universe can be removed \cite{Banados:2010ix}, but the scalar curvature constructed from the metric compatible with $\Gamma$ still blows up as we have shown. On the other hand, it is known that for the class of dark energy models with $w< -1$, i.e., the phantom models, the Big Rip singularity is inevitable for a constant $w$ in the framework of GR. Our main objective of this paper is to see if the Ba\~nados-Ferreira EiBI model can help also to remove the Big Rip singularity. We tackled this issue by investigating the possible occurrence or avoidance of doomsdays in this model. We analyzed an EiBI FLRW universe filled with dark matter and phantom energy with a constant equation of state. It is well known that a universe with such a matter-energy content under GR would face a Big Rip. Our result indicates that the Big Rip singularity remains inevitable in the EiBI theory albeit providing a minor postponement, as shown in TABLE~\ref{tabletime}. The onset of the Big Rip is independent of the amount of dark matter or dark energy; i.e. $\Omega_m$ and $\Omega_w$. In fact, the scale factor, the Hubble parameter and its cosmic time derivative all blow up in a finite cosmic time. Consequently, the scalar curvature constructed from the physical metric $g_{\mu\nu}$ will also blow up. 
We have shown as well that $R_{\mu\nu}(\Gamma)$ given in the action in Eq.(\ref{action}) and $g^{\mu\nu}R_{\mu\nu}(\Gamma)$ are infinite at the singularity, whereas $R(\Gamma)$ remains finite. The key message to take home from this letter is that a Big Rip singularity cannot be avoided in the EiBI model but it is smoother than that in GR. This is unlike the Big Loitering\footnote{We refer to this loitering effect as Big Loitering because it takes an infinite cosmic time to occur.} in a radiation dominant EiBI universe, which is rougher than that in GR, as shown in TABLE \ref{curvature}.

We will present elsewhere the behavior of other dark energy related singularities/events \cite{Nojiri:2005sx,FernandezJambrina:2006hj,BouhmadiLopez:2006fu} such as big freeze, sudden singularity, type-IV singularity, little rip, etc., in the EiBI framework \cite{progress}.

\acknowledgments

M.B.L. is supported by the Basque Foundation for Science IKERBASQUE. She also wishes to acknowledge the hospitality of LeCosPA Center at NTU (Taiwan) during the completion of part of this work and the support of the Portuguese Agency FCT through project No. PTDC/FIS/111032/2009.
C.Y.C. and P.C. are supported by Taiwan National Science Council under Project No. NSC 97-2112-M-002-026-MY3 and by Taiwan's National Center for Theoretical Sciences (NCTS). P.C. is in addition supported by US Department of Energy under Contract No. DE-AC03-76SF00515.
This work has been supported by a Spanish-Taiwanese Interchange Program with reference 2011TW0010 (Spain) and NSC 101-2923-M-002-006-MY3 (Taiwan).

\begin{table}[H]
  \begin{center}
    \begin{tabular}{||c||c|c||}
      \hline
       & Big Loitering & Big Rip \\
      \hline
      $R_{00}(g)$ &$0$ &$-\infty$ \\
      $R_{ij}(g)$ &$0$ &$+\infty$ \\
      $g^{\mu\nu}R_{\mu\nu}(g)$ &$0$ &$+\infty$ \\
      $R_{00}(\Gamma)$ &$1/\kappa$ &$-\infty$ \\
      $R_{ij}(\Gamma)$ &$-a^2\delta_{ij}/\kappa$ &$+\infty$ \\
      $g^{\mu\nu}R_{\mu\nu}(\Gamma)$ &$-4/\kappa$ &$+\infty$ \\
      $R(\Gamma)$ &$-\infty$ &$4/\kappa$ \\
      \hline
    \end{tabular}
    \caption{The behavior of different possible Ricci tensors and scalar curvatures in the EiBI model at the Big Loitering in a radiation dominated universe and at the Big Rip singularity in a phantom dominated universe.}
  \label{curvature}
  \end{center}
\end{table}

\begin{table}[H]
  \begin{center}
    \begin{tabular}{||c||c|c||}
      \hline
      $\epsilon$ & $H_0(t_{\textrm{sing}}-t_0)$(GR) &$H_0(t_{\textrm{sing}}-t_0)$(EiBI) \\
      \hline
      $0.021$ &$37.0149$ &$37.1153$ \\
      $0.041$ &$18.9291$ &$19.0294$ \\
      $0.061$ &$12.7039$ &$12.8041$ \\
      $0.081$ &$9.55371$ &$9.65379$ \\
      $0.101$ &$7.65167$ &$7.75166$ \\
      $0.121$ &$6.37884$ &$6.47876$ \\
      $0.147$ &$5.24246$ &$5.34228$ \\
      \hline
    \end{tabular}
    \caption{The cosmic time elapsed from the present time to the Big Rip singularity time, normalized to the current Hubble parameter; i.e., $H_0(t_{\textrm{sing}}-t_0)$, for different values of $\epsilon$ in GR and in the EiBI theory. We see that such cosmic time remains finite in the EiBI theory, meaning that the Big Rip singularity is inevitable. Here we assume $\Omega_m=0.287$ and $\Omega_w=0.733$, then use the constraint Eq.~(\ref{constraint}) to find the corresponding $\Omega_\kappa$.}
  \label{tabletime}
  \end{center}
\end{table}


\begin{thebibliography}{99}

\bibitem{gravitation} 
  C.~W.~Misner, K.~S.~Thorne, and J.~A.~Wheeler, \textit{Gravitation}, (W.~H.~Freeman, 1973).

\bibitem{largescale} 
  S.~W.~Hawking, G.~F.~R.~Ellis, \textit{The Large Scale Structure of Space-Time}, (Cambridge University Press, 1973).
 
\bibitem{reviewmodifiedGR}
  S.~Nojiri and S.~D.~Odintsov,
  Int.\ J.\ Geom.\ Meth.\ Mod.\ Phys.\  {\bf 4}, 115 (2007);
S.~Capozziello and M.~Francaviglia,
  Gen.\ Rel.\ Grav.\  {\bf 40}, 357 (2008);
  T.~P.~Sotiriou and V.~Faraoni,
  Rev.\ Mod.\ Phys. {\bf 82}, 451 (2010);
A.~De Felice and S.~Tsujikawa,
Living Rev.\ Rel.\  {\bf 13}, 3 (2010);
T.~Clifton, P.~G.~Ferreira, A.~Padilla and C.~Skordis,
  Phys.\ Rept.\  {\bf 513}, 1 (2012).

\bibitem{Eddington}
  A. S. Eddington, {\textit{The Mathematical Theory of Relativity}}, Cambridge University Press (1924).

\bibitem{Banados:2010ix} 
  M.~Ba\~nados and P.~G.~Ferreira,
  Phys.\ Rev.\ Lett.\  {\bf 105}, 011101 (2010).

\bibitem{Delsate:2012ky} 
  T.~Delsate and J.~Steinhoff,
  Phys.\ Rev.\ Lett.\  {\bf 109}, 021101 (2012).

\bibitem{Pani:2011mg} 
  P.~Pani, V.~Cardoso and T.~Delsate,
  Phys.\ Rev.\ Lett.\  {\bf 107}, 031101 (2011).

\bibitem{Pani:2012qb} 
  P.~Pani, T.~Delsate and V.~Cardoso,
  Phys.\ Rev.\ D {\bf 85}, 084020 (2012).

\bibitem{Avelino:2012ue} 
  P.~P.~Avelino and R.~Z.~Ferreira,
  Phys.\ Rev.\ D {\bf 86}, 041501 (2012).

\bibitem{Casanellas:2011kf} 
  J.~Casanellas, P.~Pani, I.~Lopes and V.~Cardoso,
  Astrophys.\ J.\  {\bf 745}, 15 (2012).

\bibitem{Avelino:2012ge} 
  P.~P.~Avelino,
  Phys.\ Rev.\ D {\bf 85}, 104053 (2012).

\bibitem{Avelino:2012qe} 
  P.~P.~Avelino,
  arXiv:1207.4730 [astro-ph.CO].

\bibitem{Scargill:2012kg} 
  J.~H.~C.~Scargill, M.~Ba\~nados and P.~G.~Ferreira,
Phys.\ Rev.\ D {\bf 86}, 103533 (2012).

\bibitem{EscamillaRivera:2012vz} 
  C.~Escamilla-Rivera, M.~Ba\~nados and P.~G.~Ferreira,
  Phys.\ Rev.\ D {\bf 85}, 087302 (2012).

\bibitem{Pani:2012qd} 
 P.~Pani and T.~P.~Sotiriou,
Phys.\ Rev.\ Lett.\  {\bf 109}, 251102 (2012).  

\bibitem{Starobinsky:1999yw} 
  A.~A.~Starobinsky,
  Grav.\ Cosmol.\  {\bf 6}, 157 (2000).

\bibitem{Caldwell:2003vq}
  R.~R.~Caldwell, M.~Kamionkowski and N.~N.~Weinberg,
  Phys.\ Rev.\ Lett.\  {\bf 91}, 071301 (2003).

\bibitem{Komatsu:2010fb}
  G.~Hinshaw {\it et al.},
  arXiv:1212.5226 [astro-ph.CO].


\bibitem{Nojiri:2005sx} 
  S.~'i.~Nojiri, S.~D.~Odintsov and S.~Tsujikawa,
  Phys.\ Rev.\ D {\bf 71}, 063004 (2005).

\bibitem{FernandezJambrina:2006hj}
  L.~Fern\'andez-Jambrina and R.~Lazkoz,
  Phys.\ Rev.\ D {\bf 74}, 064030 (2006).

\bibitem{BouhmadiLopez:2006fu}
  M.~Bouhmadi-L\'opez, P.~F.~Gonz\'alez-D\'iaz and P.~Mart\'in-Moruno,
  Phys.\ Lett.\ B {\bf 659}, 1 (2008).

\bibitem{progress}
  M.~Bouhmadi-L\'opez, C.~Y.~Chen, P.~Chen, work in progress.
  
\end{thebibliography}
\end{document}